\documentclass[twocolumn,preprintnumbers,amsmath,aps]{revtex4}
\usepackage[dvips]{graphicx}
\usepackage{dcolumn}
\usepackage{bm}
\usepackage{color}
\usepackage{multirow}

\begin{document}
\def\eq#1{(\ref{#1})}
\def\fig#1{Fig.\hspace{1mm}\ref{#1}}
\def\tab#1{Tab.\hspace{1mm}\ref{#1}}
%
\title{
The thermodynamic critical field and specific heat of superconducting state in phosphorene under strain  }
\author{K. A. Szewczyk$^{\left(1\right)}$}
\email{kamila.szewczyk@ajd.czest.pl}
\author{A. Z. Kaczmarek$^{\left(2\right)}$}
\author{E. A. Drzazga$^{\left(2\right)}$}

\affiliation{$^1$ Institute of Physics, Jan D{\l}ugosz University in Cz{\c{e}}stochowa, Ave. Armii Krajowej 13/15, 42-200 Cz{\c{e}}stochowa, Poland}
\affiliation{$^2$ Institute of Physics, Cz{\c{e}}stochowa University of Technology, Ave. Armii Krajowej 19, 42-200 Cz{\c{e}}stochowa, Poland}
\date{\today} 
\begin{abstract}

In this work we present the thermodynamic properties of the superconducting state in phosphorene. In particular, we have examined the electron doped ($n_{D}=1.3\times 10^{14} \rm{cm^{-2}}$) and biaxially strained (4\%) monolayer of black phosphorous, which exhibits best thermodynamic stability and highest superconducting critical temperature ($T_{c}$) among all monolayer phosphorene structures. Due to the confirmed electron-phonon pairing mechanism and relatively high electron-phonon coupling constant in the studied material, we carried out the calculations in the framework of the Eliashberg formalism for a wide range of the Coulomb pseudopotential $\mu^{\star}\in\langle 0.1, 0.3\rangle$. We have determined the thermodynamic critical field ($H_{c}$), and the specific heat difference ($\Delta C$) between superconducting ($C^{S}$) and normal state ($C^{N}$) as the functions of the temperature. In addition, we have calculated the dimensionless parameters $R_{C}=\Delta C(T_{c})/C^{N}(T_{c})$ and $R_{H}=T_{c}C^{N}(T_{c})/H^{2}_{c}(0)$, and also found their significant deviation from the expectations of the BCS theory. In particular, $R_{C} \simeq \langle 2.724, 1.899\rangle$ and $R_{H} \simeq \langle 0.133, 0.155\rangle$ for $\mu^{\star}\in \langle 0.1, 0.3\rangle$.

\noindent{\bf PACS:} 74.25.Bt, 74.20.Fg, 74.78.−w
\end{abstract}
\maketitle
\noindent{\bf Keywords:} superconductivity, thermodynamic critical field, specific heat, phosphorene

\section{Introduction}

Phosphorene - the monoatomic layer of the black phosphorous - has been paid a special attention by the world of science due to its interesting physical properties that can be used in the nano- and opto-electronics including applications in the nanoscale superconducting devices \cite{Yu2015,Dai2014,Khandelwal2017}. In 2014, it was possible to experimentally separate the one-atom layer of black phosphorus within the mechanical exfoliation \cite{Liu2014,Koenig2014} and the plasma-assisted fabrication \cite{Lu2014}. Theoretical predictions have shown that phosphorene is a semiconductor whose size of the energy gap depends on the number of material layers: 0.59 eV for 5 layers \cite{Qiao2014}  and 1.59 eV for the monolayer \cite{Rudenko2014}. These results confirm the applicability of phosphorene in the field-effect transistors  \cite{Li2014B, Ge2015}. 

The superconducting state in phosphorene, as in graphene, cannot be induced due to the zero density of states at the Fermi level. In 2014, Shao \textit{et al.} theoretically predicted that by doping with electrons $n_{D}=1.3\times 10^{14} \rm{cm^{-2}}$ the superconducting state in phosphorene can be induced, and is characterized by the critical temperature $T_{c}= 4$ K. Accordingly, it is considered that phosphorene can find application in the nanoscale superconducting devices, including superconducting quantum interference devices and superconducting transistors  \cite{Delahaye1045, Franceschi2010}.

Subsequent studies have shown that the puckered structure of phosphorene can withstand a wide range of strain \cite{Shao2014, Huang2014}, what can be used to strengthen the electron-phonon coupling and the superconducting state in phosphorene. Recent experimental data only confirmed such predictions. In an experiment investigating the superconducting state in the bilayer phosphorene \cite{Zhang2017} it was found that the critical temperature equals $T_{c}\sim3.8$ K and it is the same regardless of the intercalant alkali metal. This experiment confirms the supposition that superconductivity in phosphorene results only from its structure. In 2015, Ge \textit{et al.} have analyzed the superconducting state in phosphorene depending on the uniaxial and biaxial strain with different intensity. The increase in the critical temperature to about 16 K was shown for typical doping $n_{D}$ and the biaxial straining at a rate of 4\% \cite{Ge2015}.

Motivated by the results of the experiment and strain effects of the two-dimensional phosphorene, we have decided to examine selected thermodynamic properties of the superconducting state in the case of the biaxial straining (4\%) and in the typical doping $n_{D}=1.3\times 10^{14} \rm{cm^{-2}}$. To our knowledge, the literature data show that such phosphorene is the most stable one, and reaches the highest critical temperature among all known cases.

\section{The Eliashberg equations}

All results contained in this work have been determined basing on the Eliashberg equations in the isotropic limit for the half-filled electron band ($\langle n\rangle =1$). This is due to the relatively high electron-phonon coupling constant ($\lambda=1.6$ \cite{Ge2015}) in phosphorene. These equations take the following form \cite{Carbotte1990A, Eliashberg1960A, Marsiglio1988A, SzczesniakD2014A, Duda2015, SzczesniakK2017A, SzczesniakD2015B, SzczesniakD2015C}:
\begin{equation}
\label{r1}
\varphi_{n}=\frac{\pi}{\beta}\sum_{m=-M}^{M}
\frac{K\left(i\omega_{n}-i\omega_{m}\right)-\mu^{\star}\theta\left(\omega_{c}-|\omega_{m}|\right)}
{\sqrt{\omega_m^2Z^{2}_{m}+\varphi^{2}_{m}}}\varphi_{m}
\end{equation}
and
\begin{equation}
\label{r2}
Z_{n}=1+\frac{1}{\omega_{n}}\frac{\pi}{\beta}\sum_{m=-M}^{M}
\frac{K\left(i\omega_{n}-i\omega_{m}\right)}{\sqrt{\omega_m^2Z^{2}_{m}+\varphi^{2}_{m}}}
\omega_{m}Z_{m}.
\end{equation}
The function $\varphi_{n}\equiv \varphi(i\omega_{n})$ in the equations above is known as the order parameter, and $Z_{n}\equiv Z(i\omega_{n})$ is the wave function renormalization factor. The Matsubara frequencies are given by the following formula: $\omega_{n}\equiv \frac{\pi}{\beta}(2n-1)$, where $\beta=\frac{1}{k_{B}T}$, $k_{B}$ is the Boltzmann constant. The symbol $\theta$ in (\ref{r1}) denotes the Heaviside functions, and $\omega_{C}$ is the characteristic cut-off frequency ($\omega_{C}=3\Omega_{max}$, where $\Omega_{max}=47.332  ~\rm{meV}$).

The Eliashberg equations on the imaginary axis allow to determine the thermodynamic properties of the superconducting state in phosphorene, such as: the free energy difference ($\Delta F$) and the specific heat difference ($\Delta C$) between the superconducting state and the normal state, and the thermodynamic critical field ($H_{c}$). 

The kernel of the electron-phonon interaction takes the form:
\begin{equation}
\label{r3}
K\left(z\right)\equiv 2\int_0^{+\infty}d\omega\frac{\omega}{\omega ^2-z^{2}}\alpha^{2}F\left(\omega\right).
\end{equation}
In the considered case we assume that the function $K(z)$ has been derived in the approximation of the gas of the non-interacting phonons. The Eliashberg function ($\alpha^{2}F(\omega)$), taken from the work \cite{Ge2015}, quantitatively models the electron-phonon interaction. It was designed with the help of the Quantum Espresso package, which uses the density functional theory to determine structural, electronic, vibrational and superconducting properties of the studied material \cite{Hohenberg1964A,Giannozzi2009A}.

The depairing electron correlations are described with the help of the Coulomb pseudopotential function:

\begin{equation}
\label{r3a}
\mu^{\star}=\frac{\mu}{1+\mu\ln\left(\frac{\omega_{e}}{\omega_{ph}}\right)},
\end{equation}
where $\mu =U\rho(0) $ denotes the direct Coulomb repulsion, $U$ is the Coulomb potential, and $\rho(0)$ determines the electronic density of states on the Fermi surface.

Due to the lack of the experimental data on the superconducting state in the monolayer phosphorene, we adopted a wide range of the Coulomb pseudopotential values $\mu^{\star}=\langle 0.1,0.3\rangle$, in order to get an overview of all possible physically-relevant cases.

The Eliashberg equations have been solved for $M=1100$ Matsubara frequencies, using the methods described in works \cite{SzczesniakD2015, SzczesniakD2014, Szczesniak2012L}. 
In the case under consideration, the solutions of the Eliashberg equations are stable for $T\geq T_{0}$, where $T_{0}=2~\rm{K}$.

\section{The numerical results}

%
\begin{figure} 
\includegraphics[width=\columnwidth]{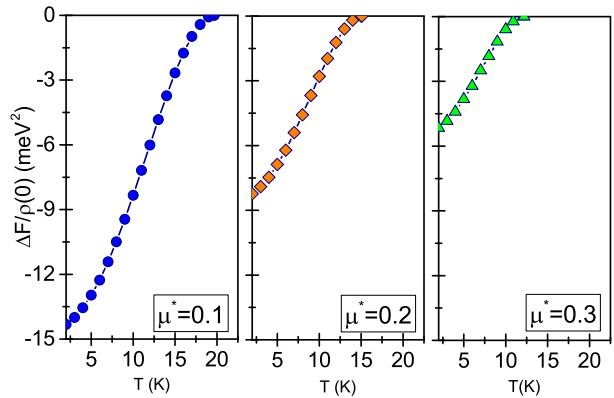}
\caption{The free energy difference between the superconducting state and normal state as a function of the temperature for three selected values of the Coulomb pseudopotential.}
\label{f2}  
\end{figure}
\begin{figure} 
\includegraphics[width=\columnwidth]{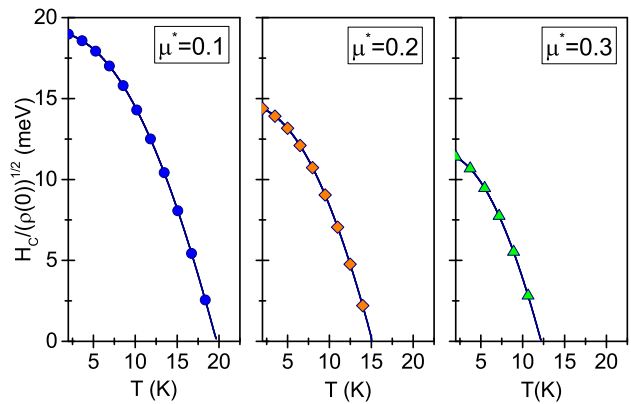}
\caption{The dependence of the thermodynamic critical field on the temperature for three selected values of the Coulomb pseudopotential.}
\label{f3}  
\end{figure}     
\begin{figure}
\includegraphics[width=\columnwidth]{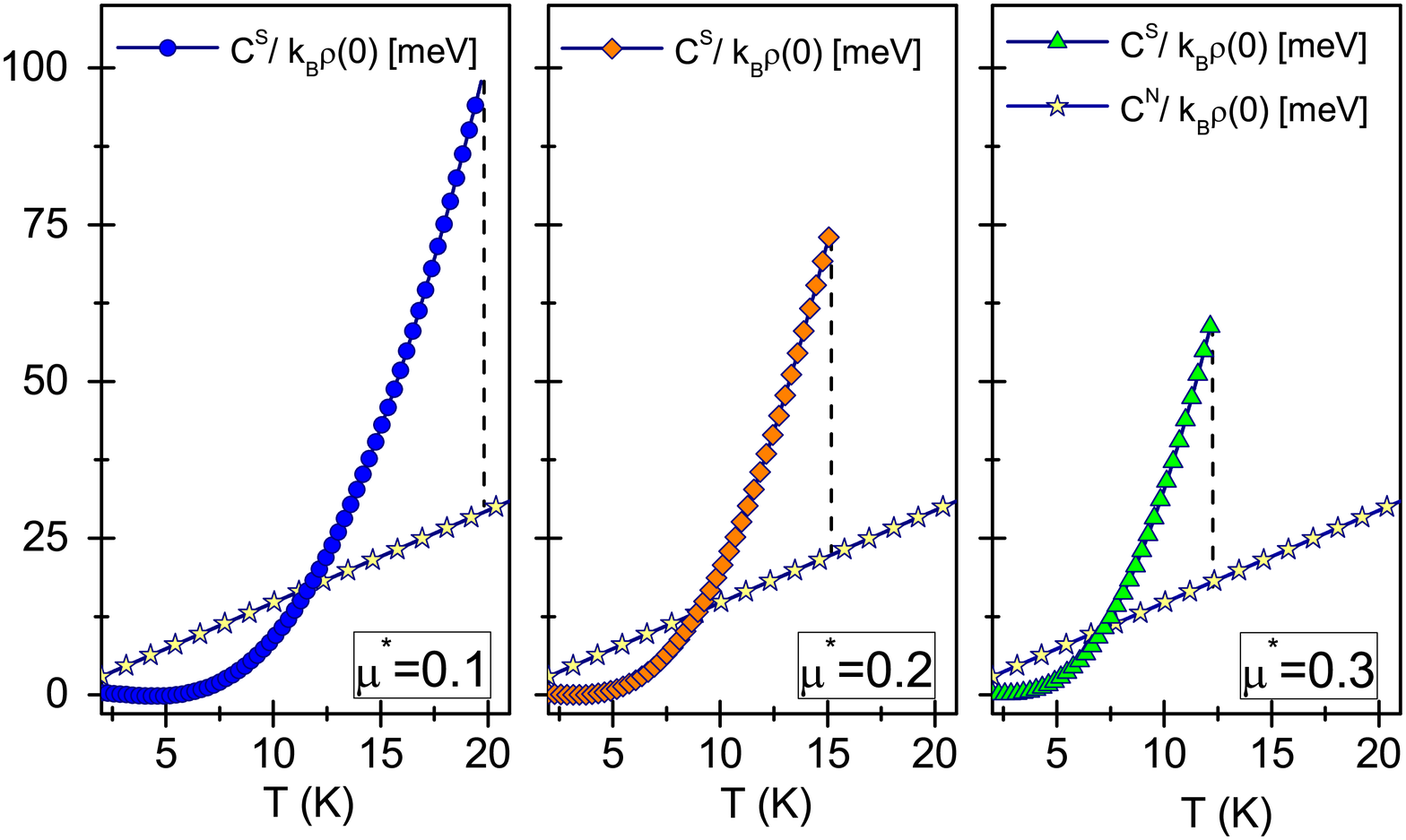}
\caption{The dependence of the specific heat of the superconducting ($\rm{C^{S}}$) and normal state ($\rm{C^{N}}$) on the temperature. The vertical lines designate characteristic jump of the specific heat for the superconducting state at $\rm{T_{c}}$.}
\label{f4}  
\end{figure}
The free energy difference between the superconducting state and the normal state ($\Delta F$), presented in figure 
\ref{f2}, was determined by applying the formula: \cite{Bardeen1964A}:
\begin{eqnarray}
\label{r4}
\frac{\Delta F}{\rho\left(0\right)} &=&-2\pi k_{B}T\sum^{M}_{n=1}
\left[\sqrt{\omega^2_n+\left(\frac{\varphi_{n}}{Z_{n}}\right)^{2}}-|\omega_n|\right]\\ \nonumber
&\times&\left[Z^{\left(S\right)}_n-Z^{\left(N\right)}_n \frac{|\omega_n|}{\sqrt{\omega^2_n+\left(\frac{\varphi_{n}}{Z_{n}}\right)^2}}\right],
\end{eqnarray}
where $Z^{\left(S\right)}_n$ and $Z^{\left(N\right)}_n$ denote the wave function renormalization factor for the superconducting state ($S$) and the normal state ($N$). The given expression (\ref{r4}) is normalized with respect to the electronic density of states on the Fermi surface. From a physical point of view, the negative values of $\Delta F/\rho\left(0\right)$ testify the thermodynamic stability of the superconducting state in phosphorene subjected to a 4\% strength strain. In addition, it is worth paying attention to the strong decline in $\Delta F(T_{0})$ in relation to $\mu^{\star}$:$[\Delta F(T_{0})]_{\mu^{\star}=0.1}/[\Delta F(T_{0})]_{\mu^{\star}=0.3}\simeq 2.77$.

\begin{figure}
\includegraphics[width=\columnwidth]{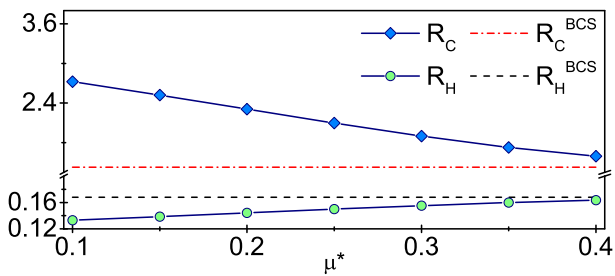}
\caption{The value of $R_{C} $ and $R_{H}$ parameters dependence on the Coulomb pseudopotential. The dashed lines indicate the parameter's value for the BCS theory. }
\label{f5}  
\end{figure}

The knowledge of the dependence (\ref{r4}) allows to estimate two consecutive thermodynamic properties of the superconducting state in the studied case \cite{Carbotte1990A}:
\begin{equation}
\label{r5}
\frac{H_{c}}{\sqrt{\rho\left(0\right)}}=\sqrt{-8\pi\left[\Delta F/\rho\left(0\right)\right]}
\end{equation}
and
\begin{equation}
\label{r6}
\frac{\Delta C\left(T\right)}{k_{B}\rho\left(0\right)}=-\frac{1}{\beta}\frac{d^{2}\left[\Delta F/\rho\left(0\right)\right]}{d\left(k_{B}T\right)^{2}}.
\end{equation}
Expression (\ref{r5}) determines the thermodynamic critical field. Its values are always positive and decrease with increasing temperature. A significant decrease in value of $H_{c}$ in relation to $\mu^{\star}$: $[H_{c}(T_{0})]_{\mu^{\star}=0.1}/[H_{c}(T_{0})]_{\mu^{\star}=0.3}\simeq 1.67$ can be seen in figure \ref{f3}.
The specific heat difference ($\Delta C=C^{S}-C^{N}$) between the superconducting state ($C^{S}$) and the normal state ($C^{N}$) is presented with the help of formula (\ref{r6}). The specific heat for the normal state can be determined using the formula: $C^{N}/ k_{B}\rho\left(0\right)=\gamma/\beta$, where $\gamma=\frac{2}{3}\pi^{2}(1+\lambda)$ is the Sommerfeld constant. Figure \ref{f4} presents the dependence of the specific heat on the temperature for three selected values of the Coulomb pseudopotential. It can be seen that $C^{S}$ for the low temperatures increases exponentially, while at higher temperatures - almost linearly. At the critical temperature it reaches the value of $C^{N}$. At the $T_{c}$ it is possible to observe a specific heat jump and the decrease in the temperature's value, at which this jump occurs relative to $\mu^{\star}$: $[\Delta C({T_{c}})]_{\mu^{\star}=0.1}/[(\Delta C({T_{c}})]_{\mu^{\star}=0.3}\simeq 1.64.$

The knowledge of the thermodynamic functions $C^{S}$, $C^{N}$ and $H_{c}$ gives the opportunity to determine the basic dimensionless parameters of the BCS theory:
\begin{equation}
\label{r7}
R_{C}=\frac{\Delta C(T_{c})}{C^{N}(T_{c})} \hspace{0.65cm} \textrm{and} \hspace{0.65cm} R_{H}=\frac{T_{c}C^{N}(T_{c})}{H^{2}_{c}(0)},
\end{equation}
where $H_{c}(0) \simeq H_{c}(T_{0})$. Their dependence on the Coulomb pseudopotential is presented in figure \ref{f5}. Dashed lines mark the values of those parameters for the BCS theory, and these values are: $R_{C}=1.43$ and $R_{H}=0.168$ \cite{Bardeen1957A,Bardeen1957B}. For phosphorene subjected to a biaxial strain we have received $R_{C} \simeq \langle 2.724, 1.899\rangle$ and $R_{H} \simeq \langle 0.133, 0.155\rangle$ for $\mu^{\star}\in \langle 0.1, 0.3\rangle$. Such a large deviation from the predictions of the BCS theory is related to the strong electron-phonon coupling and the strong retardation effects in two-dimensional phosphorene.

\section{Summary}
In the presented work we have solved the Eliashberg equations on the imaginary axis to be able to determine the selected thermodynamic properties of the superconducting state in phosphorene. In the paper, we have presented the dependencies of the free energy, the thermodynamic critical field and the specific heat on the temperature. In addition, we have presented the parameter values $R_{C}$ and $ R_{H}$ as the functions of the Coulomb pseudopotential. In particular, we have obtained $R_{C} \simeq \langle 2.724, 1.899\rangle$ and  $R_{H} \simeq \langle 0.133, 0.155\rangle$ for $\mu^{\star}\in \langle 0.1, 0.3\rangle$. This result confirms the strong-coupling character of the superconducting state in phosphorene.

%
\begin{acknowledgments}
Some calculations have been conducted on the Cz{\c{e}}stochowa University of Technology cluster, built in the framework of the
PLATON project, no. POIG.02.03.00-00-028/08 - the service of the campus calculations U3.
\end{acknowledgments}
\bibliography{template}
\end{document}